\documentclass[11pt]{article}
\usepackage{moriond}
\usepackage{epsfig}
\usepackage{amssymb}

\begin{document}

\vspace*{4 cm}

\title{GENERIC GRAVITATIONAL WAVE SIGNALS FROM THE COLLAPSE OF
  ROTATING STELLAR CORES}

\author{H.~DIMMELMEIER}
\address{Department of Physics, Aristotle University of
  Thessaloniki, \\
  GR-54124 Thessaloniki, Greece}

\author{C.~D.~OTT}
\address{Department of Astronomy and Steward Observatory,
  University of Arizona, \\
  Tucson, AZ 85721, U.S.A.}

\author{H.-T.~JANKA, A.~MAREK, E.~M\"ULLER}
\address{Max Planck Institute for Astrophysics,
  Karl-Schwarzschild-Str.\ 1, \\
  D-85741 Garching, Germany}

\maketitle\abstracts{
  We present detailed results from performing general relativistic
  (GR) simulations of stellar core collapse to a proto-neutron star,
  using a microphysical equation of state (EoS) as well as an
  approximate description of deleptonization during the collapse
  phase. We show that for a wide variety of rotation rates and
  profiles the gravitational wave (GW) burst signals from the core
  bounce are of a generic type, already known as Type~I in the
  literature. In addition, for most models the characteristic
  frequency of the GW burst signal lies in a narrow range around
  approximately $ 718 \mathrm{\ Hz} $. In our systematic study, using
  both GR and Newtonian gravity, we identify, individually quantify,
  and discuss in detail the micro- and macrophysical mechanisms
  leading to this result, i.e.\ the effects of rotation, the EoS, and
  deleptonization. We also discuss the detectability prospects of such
  GW burst signals by GW detectors, and infer that such a generic type
  of signal templates will likely facilitate a more efficient search
  in current and future detectors of both interferometric and resonant
  type.}


\section{Introduction}

Theoretical predictions of the gravitational wave (GW) signal produced
by the collapse of a rotating stellar iron core to a proto-neutron
star (PNS) in a core collapse supernova are complicated, as the
emission mechanisms are very diverse. While the prospective GW burst
signal from the collapse, bounce, and early postbounce phase is
present only when the core
rotates~\cite{mueller_82_a,moenchmeyer_91_a,zwerger_97_a,dimmelmeier_02_a,kotake_03_a,ott_04_a,ott_07_a},
GW signals with sizeable amplitudes can also be expected from
convective motions at later post-bounce phases, anisotropic
neutrino emission, excitation of various oscillations in the PNS, or
nonaxisymmetric rotational instabilities~\cite{rampp_98_a,mueller_04_a,shibata_05_a,ott_06_a,ott_07_a,ott_07_b}.

In the observational search for GWs from merging binary black holes or
neutron stars, powerful data analysis algorithms like matched
filtering are applied, as the waveform from the inspiral phase can be
modeled very accurately~\cite{blanchet_06_a}. In stark contrast, the
GW burst signal from stellar core collapse and bounce cannot yet be
predicted with the desired accuracy and robustness. First, a general
relativistic (GR) description of consistently coupled gravity and
hydrodynamics including the important microphysics is necessary. Only
very few multi-dimensional codes have recently begun to approach these
requirements. Second, the rotation rate and profile of the progenitor
core are not very strongly constrained by either observation or
numerical modeling of stellar evolution. Therefore, the influence of
rotation on the collapse dynamics and thus the GW burst signal must be
investigated by computationally expensive parameter studies.

Previous simulations, considering a large variety of rotation rates
and profiles in the progenitor core but ignoring complex (though
essential) microphysics and/or the influence of GR, found
qualitatively and quantitatively different types of GW burst signals
(see, e.g., the work by~\cite{moenchmeyer_91_a,zwerger_97_a,dimmelmeier_02_a}).
These can be classified depending on the collapse dynamics:
\emph{Type~I} signals are emitted when the collapse of the
homologously contracting inner core is not strongly influenced by
rotation, but stopped by a \emph{pressure-dominated bounce} due to the
stiffening of the EoS at nuclear density $ \rho_\mathrm{nuc} $ where the
adiabatic index $ \gamma_\mathrm{eos} $ rises above $ 4 / 3 $. This
leads to an instantaneous formation of the PNS with a maximum core
density $ \rho_\mathrm{max} \ge \rho_\mathrm{nuc} $. \emph{Type~II}
signals occur when centrifugal forces, which grow during contraction
due to angular momentum conservation, are sufficiently strong to halt
the collapse, resulting in consecutive (typically multiple)
\emph{centrifugal bounces} with intermediate coherent re-expansion of
the inner core, seen as density drops by often more than an order of
magnitude; thus here $ \rho_\mathrm{max} < \rho_\mathrm{nuc} $ after
bounce. \emph{Type~III} signals appear in a pressure-dominated bounce
when the inner core has a very small mass at bounce due to a soft
subnuclear EoS or very efficient electron capture.

In contrast, new GR simulations of rotational core collapse employing
a microphysical EoS and an approximation for deleptonization during
collapse~\cite{ott_07_a,dimmelmeier_07_a} show that the GW
burst signature is exclusively of Type~I. In a recent study, we
considerably extended the number of models and comprehensively
explored a wide parameter space of initial rotation
states~\cite{dimmelmeier_07_a}. Also for this more general
setup we found GW signals solely of Type~I form. We identified the
physical conditions that lead to the emergence of this generic GW
signal type and quantified their relative influence. These results
strongly suggest that the waveform of the GW \emph{burst} signal from
the collapse of rotating iron cores in a supernova event is much more
generic than previously thought. In this work we recapitulate the
results presented in~\cite{dimmelmeier_07_a} and discuss the
mechanisms which lead to uniformity of the signal type in rotational
supernova core collapse to a PNS in more detail.


\section{Model Setup and Numerical Methods}

We perform all simulations in 2\,+\,1 GR using the
\mbox{\textsc{CoCoNuT}}
code~\cite{dimmelmeier_02_a,dimmelmeier_05_a}, approximating GR
by the conformal flatness condition
(CFC)~\cite{isenberg_78_a,wilson_96_a}, whose excellent quality in the
context of rotational stellar core collapse has been demonstrated
extensively (see, e.g., the result presented
in~\cite{shibata_04_a,ott_07_a}). \mbox{\textsc{CoCoNuT}} utilizes
spherical coordinates with the grid setup specified in~\cite{ott_07_a}
and assumes axisymmetry. GR hydrodynamics is implemented via
finite-volume methods, piecewise parabolic reconstruction, and an
approximate Riemann solver. We use Eulerian spherical coordinates
and assume axisymmetry for the core-collapse
simulations discussed here. The computational grids consist
of 250 logarithmically-spaced and centrally-condensed radial
zones with a central resolution of $ 250 \mathrm{\ m} $
and 45 equidistant angular zones covering $ 90^\circ $.
GWs are extracted using a variant of the
Newtonian quadrupole formula~(see, e.g., the definition
in~\cite{shibata_04_a}).

We employ the microphysical EoS of Shen et al.~\cite{shen_98_a} in the
implementation of Marek et al.~\cite{marek_05_a}. Deleptonization by
electron capture onto nuclei and free protons is proposed by
Liebend\"orfer~\cite{liebendoerfer_05_a}: During collapse the electron
fraction $ Y_e $ is parameterized as a function of density based on
data from neutrino radiation-hydronamic simulations in spherical
symmetry~\cite{marek_05_a} using the latest available electron
capture rates~\cite{langanke_00_a} (updating recent
results~\cite{ott_07_a} where standard capture rates were used). After
core bounce, $ Y_e $ is only passively advected and further lepton
loss is neglected. Again following the formalism
in~\cite{liebendoerfer_05_a}, above the trapping density at
$ \rho_\mathrm{trap} = 2.0 \times 10^{12} \mathrm{\ g\ cm}^{-3} $
contributions due to neutrino radiation pressure $ P_\nu $ are taken
into account.

As initial data we take the non-rotating $ 20 \, M_\odot $
solar-metallicity progenitor s20.0 from~\cite{woosley_02_a}, imposing
the rotation law discussed in~\cite{ott_04_a,dimmelmeier_02_a}. In
order to determine the influence of different angular momentum on the
collapse dynamics, we parameterize the initial rotation of our models
in terms of the differential rotation parameter $ A $ (A1:
$ A = 50,000 \mathrm{\ km} $, almost uniform; A2: $ A = 1,000
\mathrm{\ km} $, moderately differential; A3:
$ A = 500 \mathrm{\ km} $, strongly differential) and the initial
rotation rate $ \beta_\mathrm{i} = T / |W| $, which is the ratio of
rotational energy to gravitational energy (approximately
logarithmically spaced in 18 steps from $ 0.05\% $ to $ 4\% $).

\begin{figure}[tb]
  \begin{minipage}[t]{11.0 cm}
    ~\\ [-1 em]

    \psfig{figure = 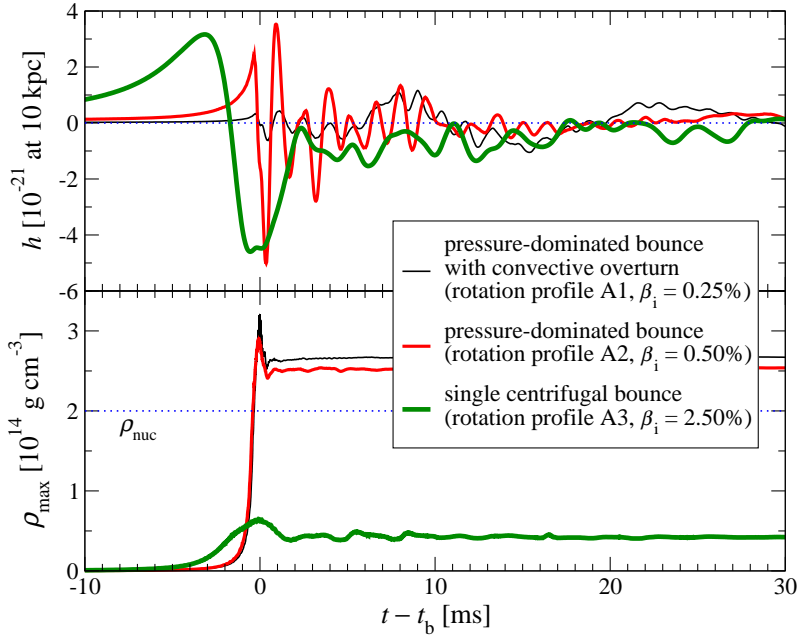, scale = 0.44}
  \end{minipage}
  \hfill
  \begin{minipage}[t]{4.5 cm}
    \caption{Time evolution of the GW amplitude $ h $ and maximum
      density $ \rho_\mathrm{max} $ for three representative models
      with different rotation profiles and initial rotation rates
      $ \beta_\mathrm{i} $. The model with slow and almost uniform
      initial rotation (black curve) develops considerable prompt
      post-bounce convection visible as a lower-frequency contribution
      in the waveform, while waveform for the model with moderate
      rotation (red curve) exhibits a regular ring-down. The maximum
      density of the rapidly rotating model which undergoes
      centrifugal bounce (green curve) remains always below nuclear
      density $ \rho_\mathrm{nuc} $. Time is nomalized to the time of
      bounce $ t_\mathrm{b} $.\hfill~
      \label{fig:generic_collapse_type}}
  \end{minipage}
\end{figure}


\section{Results}


\subsection{Generic Type of the Collapse and the Gravitational Wave Signal}

As already conjectured in~\cite{ott_07_a} and confirmed
in~\cite{dimmelmeier_07_a}, in the entire investigated parameter space
our models yield GW burst signals of Type~I, i.e.\ the waveform
exhibits a positive pre-bounce rise and then a large negative peak,
followed by a ring-down (upper panel of
Fig.~\ref{fig:generic_collapse_type}). However, with respect to
collapse dynamics and the relevant forces halting the collapse, the
models fall into two classes. While for instance all models with the
almost uniform rotation profile A1 experience a pressure-dominated
bounce for which a Type~I waveform is expected, models with profiles
A2 or A3 \emph{and} sufficiently high initial rotation rate
$ \beta_\mathrm{i} $ ($ \ge 4\% $ for A2; $\ge 1.8\% $ for A3) show a
\emph{single} centrifugal bounce at subnuclear density (lower panel of
Fig.~\ref{fig:generic_collapse_type}). Nevertheless, they also produce
a Type~I waveform, as their core does not re-expand after bounce to
densities much less than those reached at bounce but immediately
settles to a PNS after a short ring-down phase. What obviously
distinguishes models with pressure-dominated bounce from those with
centrifugal bounce is that the latter have GW signals with
significantly lower average frequencies. Note also that models with
very little rotation develop convective overturn of the shock-heated
layer immediately after shock stagnation (not to be confused with the
late-time convection discussed in~\cite{mueller_04_a}), resulting in a
lower-frequency contribution to the post-bounce GW signal (see
Fig~\ref{fig:generic_collapse_type}). These long-lasting, almost
undamped convective motions are an artifact of our insufficient
neutrino treatment \emph{after} the core bounce and are efficiently
suppressed if a more accurate description for neutrinos like Boltzmann
transport is utilized~\cite{mueller_04_a}.

\begin{figure}[tb]
  \psfig{figure = 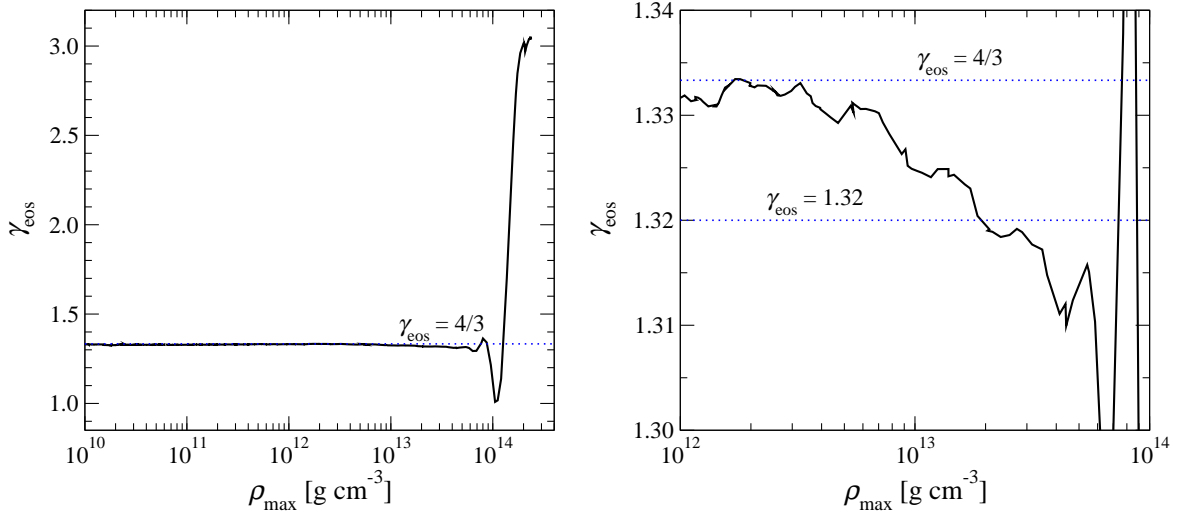, scale = 0.44}
  \caption{Left panel: Adiabatic index $ \gamma_\mathrm{eos} $ of the
    microphysical EoS in the entire density range of the maximum
    density $ \rho_\mathrm{max} $ in the core during the
    collapse of a nonrotating model. Although $ \rho_\mathrm{max} $,
    which is obtained in the center of the core, does not follow a
    trajectory of constant entropy, $ s $ is still approximately
    conserved in the pre-bounce phase. Right panel: Magnified view of
    $ \gamma_\mathrm{eos} $ in the dynamically most relevant density
    range between $ 10^{12} $ and $ 10^{14} \mathrm{\ g\ cm}^{-3} $.
    The average value of $ \gamma_\mathrm{eos} $ in this density
    regime is approximately $ 1.32 $.\hfill~
    \label{fig:gamma_eff}}
\end{figure}

In order to analyze the absence of Type~II signals, in particular for
cases with centrifugal bounce, we now separately investigate and
quantify the influence of GR, a microphysical EoS, and deleptonization
on the dynamics of rotational core collapse with different amounts and
distributions of angular momentum. When the pre-collapse iron core
starts to contract, its \emph{effective} adiabatic index
$ \gamma_\mathrm{eff} $ is lower than the critical value
$ \simeq 4 / 3 $ needed for stability against gravitational collapse.
Here $ \gamma_\mathrm{eff} $ is the sum of the adiabatic index
$ \gamma_\mathrm{eos} = \partial \ln P / \partial \ln \rho|_{Y_e, s} $
of the EoS (where $ P $ is the pressure, $ \rho $ the density, and
$ s $ the specific entropy of the fluid) and a possible correction
due to deleptonization (which can be significant until neutrino
trapping sets in at $ \rho_\mathrm{trap} $;
see~\cite{moenchmeyer_91_a}). At this stage, both GR and
rotational effects (in our range of $ \beta_\mathrm{i} $) are
negligible in discussing stability. If the build-up of centrifugal
forces in the increasingly faster spinning core during collapse is
strong enough, contraction is halted and the core undergoes a
centrifugal bounce rather than reaching nuclear density (where the
stiffening of the EoS with $ \gamma_\mathrm{eos} \gtrsim 2 \gg 4 / 3 $
would also stop the collapse; see left panel of
Fig.~\ref{fig:gamma_eff}).


\subsection{Influence of General Relativistic Gravity}

\begin{figure}[tb]
  \begin{minipage}[t]{11.0 cm}
    ~\\ [-1 em]

    \psfig{figure = 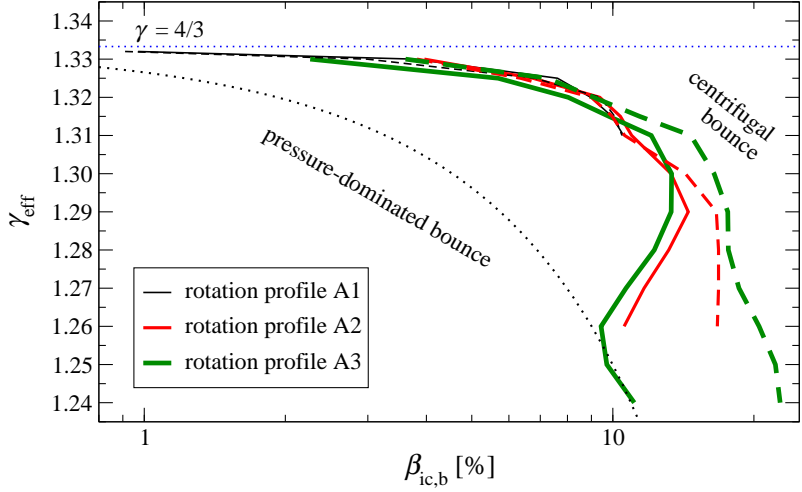, scale = 0.44}
  \end{minipage}
  \hfill
  \begin{minipage}[t]{4.5 cm}
    \caption{Boundary between pressure-dominated and centrifugal
      bounce in the $ \gamma_\mathrm{eff} $--$ \beta_\mathrm{ic,b} $
      plane for models using the hybrid EoS in Newtonian gravity
      (dashed lines) and GR (solid lines). In a dynamical core
      collapse the simple Newtonian analytic relation (curved black
      dotted line), which additionally assumes equilibrium, strongly
      underestimates the correct value for $ \beta_\mathrm{ic,b} $ in
      a wide range of $ \gamma_\mathrm{eff} $ for all investigated
      rotation profiles.\hfill~
      \label{fig:analytic_relation}}
  \end{minipage}
\end{figure}

A necessary condition for a centrifugal bounce at subnuclear densities
is that $ \gamma_\mathrm{eff} $ exceeds a critical rotation index
$ \gamma_\mathrm{rot} $. There exists a simple Newtonian analytic
relation~\cite{tohline_84_a}, $ \gamma_\mathrm{rot} =
(4 - 10 \beta_\mathrm{ic,b}) / (3 - 6 \beta_\mathrm{ic,b}) $ (where
$ \beta_\mathrm{ic,b} $ is the inner core's rotation rate at bounce),
which works well in equilibrium, but is rather imprecise as a
criterion for centrifugal bounce in a dynamical situation. For
instance, for rotating core collapse models in Newtonian gravity with
a simple hybrid EoS~\cite{janka_93_a} and no deleptonization (where
$ \gamma_\mathrm{eff} = \gamma_\mathrm{eos} $), we find that for our
range of initial rotation rates and
$ 1.24 \le \gamma_\mathrm{eff} \le 1.332 $, the analytic relation
strongly underestimates the actual $ \gamma_\mathrm{rot} $ by up to
$ \sim 0.2 $ at high $ \beta_\mathrm{ic,b} $, as shown in
Fig.~\ref{fig:analytic_relation}. Furthermore, $ \beta_\mathrm{ic,b} $
is a result of the evolution, depending on the initial parameters
$ A $ and $ \beta_\mathrm{i} $ of the pre-collapse core in an a-priori
unknown way.

\begin{figure}[tb]
  \begin{minipage}[t]{11.0 cm}
    ~\\ [-1 em]

    \psfig{figure = 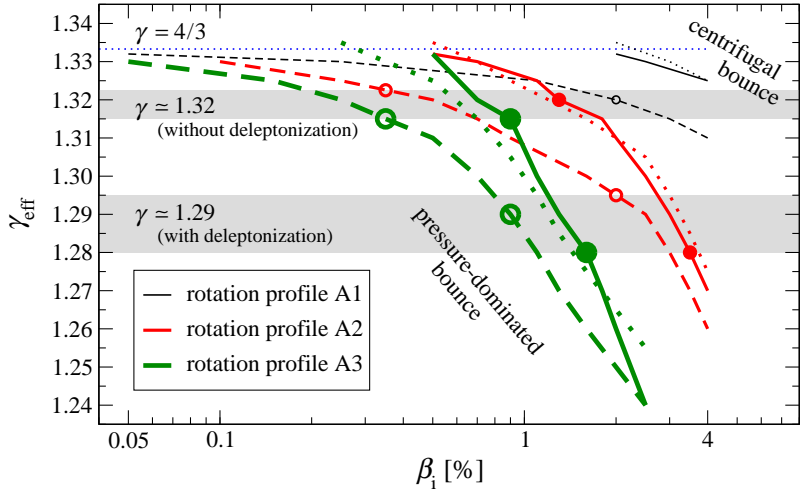, scale = 0.44}
  \end{minipage}
  \hfill
  \begin{minipage}[t]{4.5 cm}
    \caption{Boundary between pressure-dominated and centrifugal
      bounce in the $ \gamma_\mathrm{eff} $--$ \beta_\mathrm{i} $
      plane for models using the hybrid EoS in Newtonian gravity
      (dashed lines) and GR (solid lines). The curved dotted lines
      show the Newtonian results shifted by
      $ -\Delta \gamma_\mathrm{gr} = 0.015 $. The transition points
      for models using the microphysical EoS without and with
      deleptonization, again for Newtonian gravity (circles) and GR
      (bullets), lie in the shaded areas around
      $ \gamma_\mathrm{eff} \simeq 1.32 $ and $ 1.29 $,
      respectively.\hfill~
      \label{fig:relativistic_corrections}}
  \end{minipage}
\end{figure}

For this reason, for each rotation profile (specified by $ A $) we
determine the boundary between pressure-dominated and centrifugal
bounce in terms of the initial rotation rate $ \beta_\mathrm{i} $ and
the effective adiabatic index $ \gamma_\mathrm{eff} $ by systematic
numerical simulations. For models with a simple hybrid
EoS~\cite{janka_93_a} and no deleptonization, using the same initial
density profile as in the microphysical models, the results are shown
in Fig.~\ref{fig:relativistic_corrections}, both in the Newtonian case
(dashed lines) and in GR (solid lines). As is apparent, for our choice
of initial rotation the influence of GR can be approximated by adding
an offset of $ -\Delta \gamma_\mathrm{gr} \simeq 0.015 $ to the
Newtonian results (dotted lines). This gives a quantitative measure of
the GR effects on rotational core collapse, which is in agreement
with~\cite{dimmelmeier_02_a}. Note that $ \Delta \gamma_\mathrm{gr} $
is negative because GR effectively acts like a softening of the EoS.

\begin{figure}[tb]
  \psfig{figure = 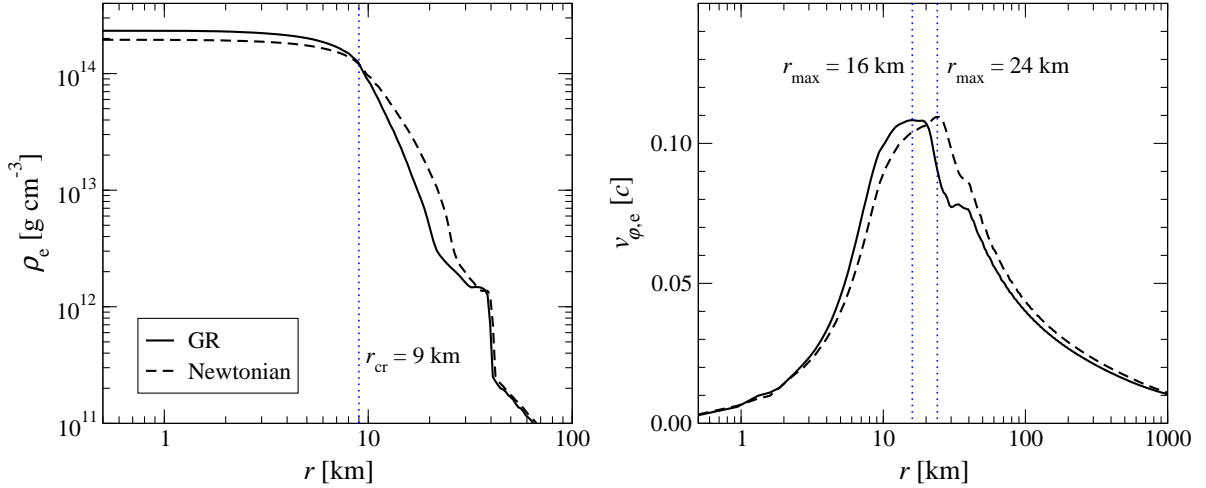, scale = 0.44}
  \caption{Left panel: Radial profile of the density
    $ \rho_\mathrm{e} $ in the equatorial plane shortly after core
    bounce for the microphysical model with rotation profile A1 and
    initial rotation rate $ \beta_\mathrm{i} = 1.10\% $ in Newtonian
    gravity (dashed line) and GR (solid line). In the central region at
    radii smaller than the density crossing radius
    $ r_\mathrm{cr} = 9 \mathrm{\ km} $ the density is higher in
    GR. Right panel: Radial profiles of the rotation velocity
    $ v_{\varphi,\mathrm{e}} $ in the equatorial plane shortly after
    core bounce for the same model. The vertical lines indicate the
    radius $ r_\mathrm{max} $ of the rotation velocity maximum.\hfill~
    \label{fig:gr_comparison}}
\end{figure}

For all 54 microphysical models, we find that the maximum density
$ \rho_\mathrm{max} $ in the core during and after bounce is always
higher in GR than in Newtonian gravity, an observation that is well
known from models with the simple hybrid
EoS~\cite{dimmelmeier_02_a}. In the nonrotating limit, the difference
in $ \rho_\mathrm{max} $ is roughly $ 10\% $, growing strongly with
increasing rotation (obviously in particular for cases where GR
produces a regular bounce while Newtonian gravity results in a
centrifugal bounce). As a consequence, the density crossing phenomenon
discussed in detail for models with a hybrid EoS
in~\cite{dimmelmeier_02_a} also occurs if microphysics is taken into
account (see left panel of Fig.~\ref{fig:gr_comparison}). In the
central parts of the PNS, the effectively stronger gravitational pull
of GR results in a higher density, while the core exhibits a lower
density compared to a Newtonian simulation outside a density crossing
radius $ r_\mathrm{cr} $. Thus the PNS is more compact in GR.

However, in contrast to previous simulations using the hybrid
EoS, for the microphysical models we do not find a clear indication
that the influence of GR consistently leads to higher infall
velocities $ v_r $ in the contraction phase and higher rotation
velocities $ v_\varphi $ during and after core bounce. While our
simulations confirm the results using the simpler models
in~\cite{dimmelmeier_02_a} that the higher compactness of the PNS in
GR translates to a smaller radius for the maximum of $ v_\varphi $, as
shown in Fig.~\ref{fig:gr_comparison}, the differences in the maximum
values for both $ v_r $ and $ v_\varphi $ are too small to be
significant. This already indicates that the effect of the GR
correction $ -\Delta \gamma_\mathrm{gr} \simeq 0.015 $ is smaller than
the influence due to microphysics, which we investigate in the
following.


\subsection{Influence of the Microphysical Equation of State}

Fig.~\ref{fig:relativistic_corrections} also shows for each rotation
profile the locations where the transition between pressure-supported
and centrifugal bounce occurs when the microphysical EoS is used and
$ \beta_\mathrm{i} $ is gradually increased from $ 0.05\% $ to
$ 4\% $. These transitions are marked on the different boundary lines
and allow the identification of the $ \gamma_\mathrm{eff} $ value
where models with the hybrid EoS make this transition. For simulations
with microphysical EoS but no deleptonization we find that in all
cases the transition occurs near $ \gamma_\mathrm{eff} \simeq 1.32 $
(highlighted by the upper grey band in
Fig.~\ref{fig:relativistic_corrections}). This value agrees with the
average of $ \gamma_\mathrm{eos} $ for the microphysical EoS at
densities between $ 10^{12} $ and $ 10^{14} \mathrm{\ g\ cm}^{-3} $
(see right panel of Fig.~\ref{fig:gamma_eff}), which is the most
relevant range for the collapse dynamics. Thus the type of core bounce
obtained with the microphysical EoS is well reproduced by the simple
hybrid EoS (which is identical to a polytrope before core bounce) with
$ \gamma_\mathrm{eos} \simeq 1.32 $.


\subsection{Influence of Deleptonization and Suppression of Type II
  Collapse}

Deleptonization before neutrino trapping reduces
$ \gamma_\mathrm{eff} $ compared to $ \gamma_\mathrm{eos} $ locally
according to $ \Delta \gamma_e = \frac{4}{3} \, \delta \ln Y_e /
\delta \ln \rho|_m < 0 $ (along trajectories of a collapsing fluid
element $ m $; see~\cite{van_riper_81_a,moenchmeyer_91_a}), resulting
in an effective softening of the EoS. Above trapping density at
$ \rho_\mathrm{trap} $ an additional positive correction
$ \Delta \gamma_\nu \approx \delta (P_\nu / P) / \delta \ln \rho|_m $
(assuming that $ P_\nu << P $) due to neutrino radiation pressure
effects must be considered. From the $ Y_e $--$ \rho $ trajectories
used to describe the deleptonization during core collapse,
$ \Delta \gamma_e $ amounts to about $ -0.06 $ to $ -0.05 $, while a
simple analytic estimate for $ \Delta \gamma_\nu $ yields roughly
$ 0.03 $. We thus anticipate values between $ -0.03 $ and $ -0.02 $
for the sum $ \Delta \gamma_e + \Delta \gamma_\nu $, again in the
density regime relevant for the bounce dynamics. Adding this
correction to $ \gamma_\mathrm{eos} \simeq 1.32 $ we expect an
effective adiabatic index $ \gamma_\mathrm{eff} \approx 1.29 $ for
models with microphysical EoS \emph{and} deleptonization. Again this
cumulative value agrees with the results obtained in our
simulations. Fig.~\ref{fig:relativistic_corrections} shows that the
bullets and circles marking those models on the different boundary
lines (for the investigated initial rotation profiles with either
Newtonian gravity or GR) all lie in the range of values indicated by
the lower grey band around $ \gamma_\mathrm{eff} \simeq 1.29 $.

The finding that deleptonization decreases $ \gamma_\mathrm{eff} $ to
about $ 1.29 $ explains the absence of Type~II GW signals for all our
models in GR with microphysics. When a hybrid EoS is used, the
subgroup of such models showing multiple centrifugal bounces and
subsequent strong re-expansion phases of the inner core occupies only
a small area in the $ \gamma_\mathrm{eff} $--$ \beta_\mathrm{i} $
plane. It is located at $ \gamma_\mathrm{eff} \ge 1.31 $ for all of
our initial rotation states both in the Newtonian case and in GR,
i.e.\ significantly above the value of
$ \gamma_\mathrm{eff} \simeq 1.29 $ that characterizes the
microphysical models if deleptonization is included.


\subsection{Mass of the Inner Core at the Time of Bounce and
  Suppression of Type III Collapse}

\begin{figure}[tb]
  \begin{minipage}[t]{10.0 cm}
    ~\\ [-1 em]

    \psfig{figure = 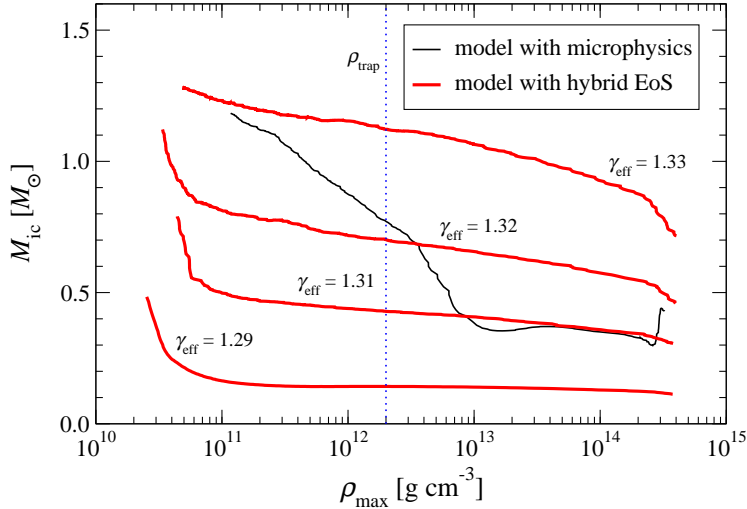, scale = 0.44}
  \end{minipage}
  \hfill
  \begin{minipage}[t]{5.5 cm}
    \caption{Mass $ M_\mathrm{ic} $ of the homologously collapsing
      inner core in the entire density range of the maximum density
      $ \rho_\mathrm{max} $ in the core during a nonrotating
      collapse in GR. The value of $ M_\mathrm{ic} $ at the high
      density end point of the curves corresponds to the mass
      $ M_\mathrm{ic,b} $ of the inner core at bounce. While the value
      of $ M_\mathrm{ic} $ for models with a simple hybrid EoS is
      already approximately determined at low densities (red curves),
      in the microphysical model it decreases considerably around
      neutrino trapping density $ \rho_\mathrm{trap} $ (black curve),
      albeit only to a value that is significantly higher than if the
      hybrid EoS with $ \gamma_\mathrm{eff} = 1.29 $ is used.\hfill~
      \label{fig:inner_core_mass}}
  \end{minipage}
\end{figure}

The value $ \gamma_\mathrm{eff} = 1.29 $ captures well the
deleptonization effects in the density regime between $ 10^{12} $ and
$ 10^{14} \mathrm{\ g\ cm}^{-3} $ (i.e.\ above neutrino trapping).
Therefore it serves as a good criterion to determine the high density
collapse dynamics and also the core bounce behavior. Consequently,
models with hybrid EoS and $ \gamma_\mathrm{eos} = 1.29 $ exhibit the
same collapse and bounce behavior in this dynamical
phase. Nevertheless, the global assumption
$ \gamma_\mathrm{eff} = 1.29 $ fails to correctly predict the mass
$ M_\mathrm{ic,b} $ of the inner core at bounce for the microphysical
models. For this value of the effective adiabatic index, by equating
the pressures of two polytropes
$ P = K (Y_{e,\mathrm{i}} = 0.5) \, \rho^{\gamma_\mathrm{eff}} $ and
$ P = K (Y_e)\,  \rho^{4 / 3} $ with $ K = K' \, Y_e^{4 / 3} $ and
constant $ K' = 1.2435 \times 10^{15} $ (in cgs units) for a
relativistic degenerate electron gas, the hybrid EoS yields an
effective average $ Y_e $ of $ \approx 0.237 $ assuming a typical mean
density in the core of $ \rho = 10^{10} \mathrm{\ g\ cm}^{-3} $ during
collapse. Using this estimate for models with the hybrid EoS we find a
small mass $ M_\mathrm{ic} \sim 0.1 \mbox{\,--\,} 0.3 \, M_\odot $
(higher for more rapid rotation), consistent with the theory of
self-similar collapse~\cite{yahil_83_a}. Due to the instantaneous
initial pressure reduction throughout the entire core in the case of
the hybrid EoS, the final value $ M_\mathrm{ic,b} $ at the time of
bounce is already determined at low densities (see
Fig.~\ref{fig:inner_core_mass}, where we plot the change of $
M_\mathrm{ic} $ with the maximum density $ \rho_\mathrm{max} $ in the
core during collapse for models without rotation).

In models with microphysics however, in the early collapse phase at
low densities the effective adiabatic index $ \gamma_\mathrm{eff} $ is
significantly higher than $ 1.29 $, both because the adiabatic index
$ \gamma_\mathrm{eos} $ of the EoS is much closer to $ 4 / 3 $ (see
Fig.~\ref{fig:gamma_eff}) and because deleptonization is weak at those
densities. This results in a high initial value
$ M_\mathrm{ic} \simeq 1.2 \, M_\odot $ for the mass of the inner core
(see Fig.~\ref{fig:inner_core_mass}). At intermediate densities
\emph{around} the neutrino trapping density $ \rho_\mathrm{trap} $,
deleptonization indeed reduces $ M_\mathrm{ic} $, but only to about
$ 0.5 \mbox{\,--\,} 0.9 \, M_\odot $ (increasing with rotation). In
the late, high density phase of the collapse, $ M_\mathrm{ic} $ then
stays rougly constant until core bounce. This behavior is in agreement
with recent spherically symmetric GR results using Boltzmann neutrino
transport~\cite{hix_03_a}.

Therefore, even by taking into account the \emph{local} influence of
deleptonization the microphysical models end up with a significantly
larger mass $ M_\mathrm{ic,b} $ of the inner core at bounce than the
models with a simple hybrid EoS with $ \gamma_\mathrm{eff} $ which
undergo a \emph{global} initial reduction of pressure. This explains
the complete absence of rapid collapse dynamics and the according
Type~III GW burst signals in our models, which only occurs for
$ M_\mathrm{ic} \lesssim 0.2 \, M_\odot $ (see also the discussion
in~\cite{kotake_03_a}).


\subsection{Detectability of the Gravitational Wave Signal}

\begin{figure}[tb]
  \begin{minipage}[t]{11.0 cm}
    ~\\ [-1 em]

    \psfig{figure = 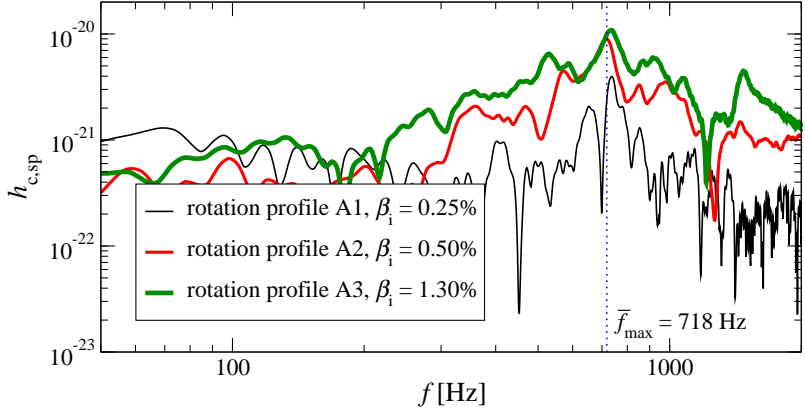, scale = 0.44}
  \end{minipage}
  \hfill
  \begin{minipage}[t]{4.5 cm}
    \caption{Characteristic GW strain spectra $ h_\mathrm{c,sp} $ at a
      distance $ d = 10 \mathrm{\ kpc} $ to the source for three
      representative models in GR with microphysical EoS and
      deleptonization that do not undergo centrifugal
      bounce. As for most other models the individual maxima
      $ f_\mathrm{max} $ of their frequency spectrum is very close to
      $ \bar{f}_\mathrm{max} \simeq 718 \mathrm{\ Hz} $.\hfill~
      \label{fig:spectrum}}
  \end{minipage}
\end{figure}

\begin{figure}[tb]
  \begin{minipage}[t]{11.0 cm}
    ~\\ [-1 em]

    \psfig{figure = 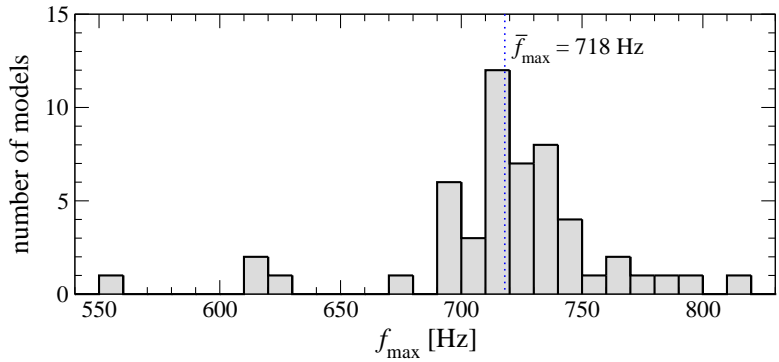, scale = 0.44}
  \end{minipage}
  \hfill
  \begin{minipage}[t]{4.5 cm}
    \caption{Histogram of the maxima $ f_\mathrm{max} $ of the
      waveform's frequency spectrum for all 54 GR models
      with microphysical EoS and deleptonization. The values of
      $ f_\mathrm{max} $ for most models are spread in a very narrow
      frequency range around the average value located at
      $ \bar{f}_\mathrm{max} \simeq 718 \mathrm{\ Hz} $.\hfill~
      \label{fig:frequency_histogram}}
  \end{minipage}
\end{figure}

The generic nature of the GW burst signal has several important
implications for prospective detectability. The limitation to a unique
Type~I waveform for a very broad range of rotation states of the
progenitor core will very likely facilitate the use of more powerful
and finetuned data analysis methods in GW detectors. To this end, we
offer our results in a publicly accessible waveform
catalog~\footnote{www.mpa-garching.mpg.de/rel\_hydro/wave\_catalog.shtml.}.
Note that almost all investigated models result in a
pressure-dominated bounce and instantaneous formation of a PNS with
similar average density and compactness. For these models, whose
rotation rates at bounce span two orders of magnitude
($ 0.2\% \lesssim \beta_\mathrm{ic,b} \lesssim 20\% $), the individual
maxima $ f_\mathrm{max} $ of their waveform's frequency spectrum lie
for most models in a very narrow range with an average of
$ \bar{f}_\mathrm{max} \simeq 718 \mathrm{\ Hz} $. This is exemplified
in Fig.~\ref{fig:spectrum}, where the characteristic GW strain
spectrum $ h_\mathrm{char} = d^{-1} \sqrt{2\pi^{-2} dE_\mathrm{GW} / df} $
(with $ E_\mathrm{gw} $ being the energy emitted in
GWs and $ d = 10 \mathrm{\ kpc} $ being the distance to the
source)~\cite{flanagan_98_a,dimmelmeier_05_a} for three representative
models that do not undergo centrifugal bounce is plotted.

\begin{figure}[tb]
  \begin{minipage}[t]{11.7 cm}
    ~\\ [-1 em]

    \psfig{figure = 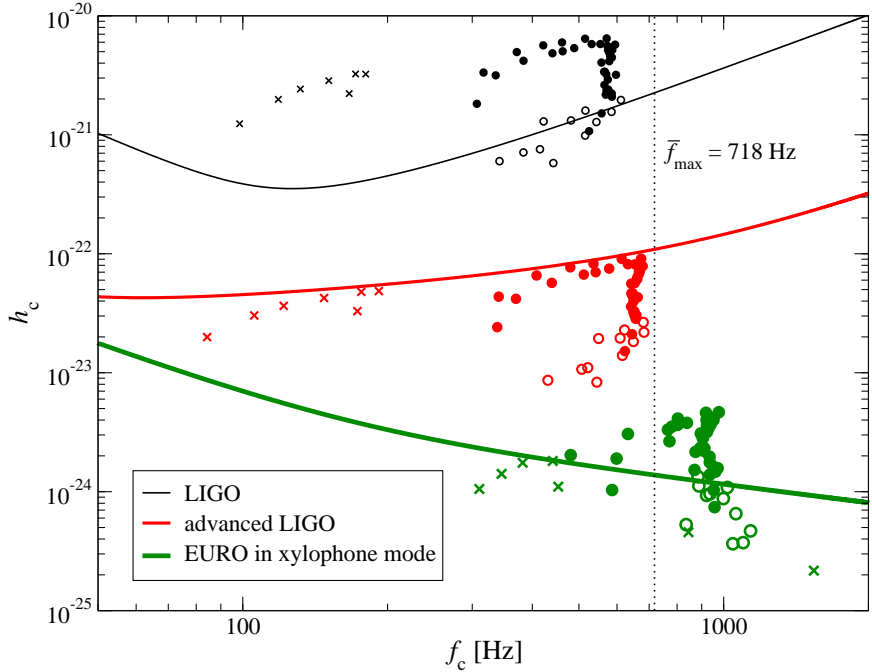, scale = 0.44}
  \end{minipage}
  \hfill
  \begin{minipage}[t]{3.8 cm}
    \caption{Location of the GW burst signals from the core bounce for
      all models in the $ h_\mathrm{c} $--$ f_\mathrm{c} $ plane
      relative to the sensitivity curves of various GW interferometer
      detectors (as color-coded). The sources are at a distance
      of $ 10 \mathrm{\ kpc} $ for LIGO, $ 0.8 \mathrm{\ Mpc} $ for
      advanced LIGO, and $ 15 \mathrm{\ Mpc} $ for EURO. Open circles
      denote models where the filtered-out early postbounce convection
      contributes significantly to the original signal, and crosses
      show models undergoing centrifugal bounce.\hfill~
      \label{fig:signal_detectability}}
  \end{minipage}
\end{figure}

The clustering in frequency can also be clearly seen in the histogram
in Fig.~\ref{fig:frequency_histogram}, where all 54 GR models
with microphysical EoS and deleptonization are shown. This property of
the GW burst signal could potentially become important for a possible
detectability by detectors of both interferometric and resonant type.
In Fig.~\ref{fig:signal_detectability} we plot the
(detector-dependent) frequency-integrated characteristic waveform
amplitude $ h_\mathrm{c} $ against the characteristic frequency
$ f_\mathrm{c} $ (Eq.~(31) in~\cite{thorne_87_a}) for all 54 GR models
with microphysical EoS and deleptonization. We assume optimal
orientation of source and detector, and in cases with
pressure-dominated bounce remove the lower-frequency contribution from
the post-bounce convective overturn by cutting the spectrum below
$ 250 \mathrm{\ Hz} $, as we are only interested in the GW signal from
the bounce and ring-down. Fig.~\ref{fig:signal_detectability} shows
that while current LIGO class interferometric detectors are only
sensitive to signals coming from an event in the Milky Way, advanced
LIGO could marginally detect some signals from other galaxies in the
Local Group like Andromeda. For the proposed EURO
detector~\footnote{www.astro.cardiff.ac.uk/geo/euro/.} in xylophone
mode, we expect a very high signal-to-noise ratio (which is
$ h_\mathrm{c} $ divided by the detector sensitivity at
$ f_\mathrm{c} $). This detector could also measure many of the
computed signals at a distance of $ 15 \mathrm{\ Mpc} $, i.e.\ in the
Virgo cluster.

Note that detectability could be enhanced by (i) a network of
interferometers in coincidence search, (ii) the support by resonant
detectors, which is particularly facilitated by the narrow range of
$ f_\mathrm{max} $, and (iii) the use of more powerful data analysis
methods beyond time-frequency analysis (see~\cite{arnaud_03_a} and
references therein) based on the waveforms' similarity and
robustness. A serious obstacle for detection is the low event rate of
$ \sim 1 \mathrm{\ yr}^{-1} $ within $ 15 \mathrm{\ Mpc} $ (or at most
$ \sim 5 \mathrm{\ yr}^{-1} $ including the entire Virgo
cluster~\cite{arnaud_04_a,nutzman_04_a,izzard_04_a}), which is further
reduced by assuming that only a small fraction (possibly only $ 1\% $)
of all progenitors rotate fast enough to have a strong GW bounce
signal~\cite{woosley_06_a}. Nevertheless, our results can serve as a
guideline for a possible frequency narrow-banding of future
interferometers (which can significantly boost sensitivity), as well
as for choosing the optimal configuration in the planning of resonant
detectors like the proposed DUAL detector~\cite{cerdonio_01_a}. In
addition, as a core collapse may be accompanied by other GW emission
mechanisms of comparable strength like late-time convection (also in a
nonrotating core), PNS pulsations, or bar-mode instabilities, the
\emph{total} GW signal strength and duration could be significantly
higher than predicted here.

As a downside, the generic properties of the GW burst signal introduce
a frequency degeneracy into the signal inversion problem.
Consequently, in the case of a detection it is difficult to
extract details about the rotation state of the pre-collapse core, because a
large part of the corresponding parameter space yields a
pressure-dominated bounce and thus signals with very similar values
for $ f_\mathrm{max} $ and (depending on the detector) also for
$ f_\mathrm{c} $. On the other hand, as $ f_\mathrm{max} $ directly
depends on the compressibility of the nuclear EoS at bounce, 
determining this frequency from the GW burst signal can help to
constrain the EoS properties around nuclear density.


\section*{Acknowledgments}

We thank D.~Shoemaker for helpful discussions. This work was supported
by DFG (SFB/TR~7 and SFB~375). H.D.\ is a Marie Curie Intra-European
Fellow within the 6th European Community Framework Programme (IEF
040464), and C.D.O.\ acknowledges support by the Joint Institute for
Nuclear Astrophysics (JINA) under NSF grant PHY0216783.



\section*{References}

\begin{thebibliography}{99}

\bibitem{mueller_82_a}
  E.~M\"uller,
  {\it Astron. Astrophys.} {\bf 114}, 53 (1982).

\bibitem{moenchmeyer_91_a}
  R.~M\"onchmeyer {\it et al},
  {\it Astron. Astrophys.} {\bf 246}, 417 (1991).

\bibitem{zwerger_97_a}
  T.~Zwerger and E.~M\"uller,
  {\it Astron. Astrophys.} {\bf 320}, 209 (1997).

\bibitem{dimmelmeier_02_a}
  H.~Dimmelmeier, J.~Font, and E.~M\"uller,
  {\it Astron. Astrophys.} {\bf 393}, 523 (2002).

\bibitem{kotake_03_a}
  K.~Kotake, S.~Yamada, and K.~Sato,
  {\it Phys. Rev.} D {\bf 68}, 044023 (2003).

\bibitem{ott_04_a}
  C.~D.~Ott {\it et al},
  {\it Astrophys. J.} {\bf 600}, 834 (2004).

\bibitem{ott_07_a}
  C.~D.~Ott {\it et al},
  {\it Phys. Rev. Lett.}, accepted (2007).

\bibitem{dimmelmeier_07_a}
  H.~Dimmelmeier {\it et al},
  {\it Phys. Rev. Lett.}, accepted (2007).

\bibitem{rampp_98_a}
  M.~Rampp, E.~M\"uller, and M.~Ruffert,
  {\it Astron. Astrophys.} {\bf 332}, 969 (1998).

\bibitem{mueller_04_a}
  E.~M\"uller {\it et al},
  {\it Astrophys. J.} {\bf 603}, 221 (2004).

\bibitem{shibata_05_a}
  M.~Shibata and Y.-I.~Sekiguchi,
  {\it Phys. Rev.} D {\bf 71}, 024014 (2005).

\bibitem{ott_06_a}
  C.~D.~Ott {\it et al},
  {\it Phys. Rev. Lett.} {\bf 96}, 201102 (2006).

\bibitem{ott_07_b}
  C.~D.~Ott {\it et al},
  {\it Class. Quantum Grav.}, accepted (2007).

\bibitem{blanchet_06_a}
  L.~Blanchet,
  {\it Living Rev. Relativity}, {\bf 9}, 4 (2006),  URL (cited on
  15.05.2007): \\
  http://www.livingreviews.org/lrr-2006-4.

\bibitem{dimmelmeier_05_a}
  H.~Dimmelmeier {\it et al},
  {\it Phys. Rev.} D {\bf 71}, 064023 (2005).

\bibitem{isenberg_78_a}
  J.~A.~Isenberg,
  University of Maryland Preprint (1978),
  gr-qc/0702113

\bibitem{wilson_96_a}
  J.~R.~Wilson, G.~J.~Mathews, and P.~Marronetti,
  {\it Phys. Rev.} D {\bf 54}, 1317 (1996).

\bibitem{shibata_04_a}
  M.~Shibata and Y.-I.~Sekiguchi,
  {\it Phys. Rev.} D {\bf 69}, 084024 (2004).

\bibitem{shen_98_a}
  H.~Shen {\it et al},
  {\it Prog. Theor. Phys.} {\bf 100}, 1013 (1998).

\bibitem{marek_05_a}
  A.~Marek {\it et al},
  {\it Astron. Astrophys.} {\bf 443}, 201 (2005).

\bibitem{liebendoerfer_05_a}
  M.~Liebend\"orfer,
  {\it Astrophys. J.} {\bf 633}, 1042 (2005).

\bibitem{langanke_00_a}
  K.~Langanke and G.~Mart\'{\i}nez-Pinedo,
  {\it Nucl. Phys.} A {\bf 673}, 481 (2000).

\bibitem{woosley_02_a}
  S.~E.~Woosley, A.~Heger, and T.~A.~Weaver,
  {\it Rev. Mod. Phys.} {\bf 74}, 1015 (2002).

\bibitem{tohline_84_a}
  J.~E.~Tohline,
  {\it Astrophys. J.} {\bf 285}, 721 (1984).

\bibitem{janka_93_a}
  H.-T.~Janka, T.~Zwerger, and R.~M\"onchmeyer,
  {\it Astron. Astrophys.} {\bf 268}, 360 (1993).

\bibitem{van_riper_81_a}
  K.~A.~van~Riper and J.~M.~Lattimer,
  {\it Astrophys. J.} {\bf 249}, 270 (1981).
        
\bibitem{yahil_83_a}
  A.~Yahil, 
  {\it Astrophys. J.} \textbf{265}, 1047 (1983).
  
\bibitem{hix_03_a}
  W.~R.~Hix {\it et al},
  {\it Phys. Rev. Lett.} {\bf 91}, 201102 (2003).

\bibitem{flanagan_98_a}
  \'E~\'E~Flanagan and S.~A.~Hughes,
  {\it Phys. Rev.} D {\bf 57}, 4535 (1998).
  
\bibitem{thorne_87_a}
  K.~S.~Thorne, in
  {\it 300 Years of Gravitation},
  ed.\ S.~W. Hawking and W.~Israel
  (Cambridge University Press, Cambridge, UK, 1987).
  
\bibitem{arnaud_03_a}
  N.~Arnaud {\it et al},
  {\it Phys. Rev.} D {\bf 67}, 062004 (2003).

\bibitem{arnaud_04_a}
  N.~Arnaud {\it et al},
  {\it Astropart. Phys.} {\bf 21}, 201 (2004).

\bibitem{nutzman_04_a}
  P.~Nutzman {\it et al},
  {\it Astrophys. J.} {\bf 612}, 364 (2004).
  
\bibitem{izzard_04_a}
  R.~G.~Izzard, E.~Ramirez-Ruiz, and C.~A.~Tout,
  {\it Mon. Not. R. Astron. Soc.} {\bf 348}, 1215 (2004).

\bibitem{woosley_06_a}  
  S.~E.~Woosley and A.~Heger,
  {\it Astrophys. J.} {\bf 637}, 914 (2006).

\bibitem{cerdonio_01_a}
  M.~Cerdonio {\it et al},
  {\it Phys. Rev. Lett.} {\bf 87}, 031101 (2001).
  
\end{thebibliography}
\end{document}